\documentclass[twoside,11pt]{article}

\usepackage{jsat}
\usepackage{pst-tree}
\usepackage{color}
\usepackage{tabularx}
\usepackage{epsfig}
\usepackage{amsmath}
\usepackage{amssymb}
\usepackage{graphicx}
\usepackage{wasysym}
\usepackage{comment}
\usepackage{float}
\usepackage{algorithm}
\usepackage[noend]{algpseudocode}
\usepackage{cleveref}

\input epsf

\definecolor{myblue}{rgb}{0.06, 0.3, 0.57}
\newcommand{\bor}{{\em \color{myblue} borealis}}

\ShortHeadings{A generalized global update algorithm for Boolean 
optimization problems}
{Z.~Zhu et al.}

\begin{document}

\title{\bor~--- A generalized global update algorithm for \, 
Boolean optimization problems}

\author{ \name \!\!Zheng Zhu
	 \email zzwtgts@tamu.edu\\
	 \name Chao Fang
	 \email fangchao118@physics.tamu.edu\\
	 \addr Department of Physics and Astronomy, \\
	       Texas A\&M University, College Station, Texas 77843-4242, USA
	 \AND
         \name  Helmut G.~Katzgraber
	 \email hgk@tamu.edu\\
         \addr  Department of Physics and Astronomy, \\
                Texas A\&M University, College Station, Texas 77843-4242,USA\\
		Santa Fe Institute, 
		1399 Hyde Park Road, Santa Fe, New Mexico 87501 USA
	 }
     
\maketitle

\begin{abstract}

Optimization problems with Boolean variables that fall into the
nondeterministic polynomial (NP) class are of fundamental importance in
computer science, mathematics, physics and industrial applications. Most
notably, solving constraint-satisfaction problems, which are related to
spin-glass-like Hamiltonians in physics, remains a difficult numerical
task. As such, there has been great interest in designing efficient
heuristics to solve these computationally difficult problems. Inspired
by parallel tempering Monte Carlo in conjunction with the rejection-free
isoenergetic cluster algorithm developed for Ising spin glasses, we
present a generalized global update optimization heuristic that can be
applied to different NP-complete problems with Boolean variables. The
global cluster updates allow for a wide-spread sampling of phase space,
thus considerably speeding up optimization. By carefully tuning the
pseudo-temperature (needed to randomize the configurations) of the
problem, we show that the method can efficiently tackle optimization
problems with over-constraints or on topologies with a large
site-percolation threshold. We illustrate the efficiency of the
heuristic on paradigmatic optimization problems, such as the maximum
satisfiability problem and the vertex cover problem.

\end{abstract}



\section{Introduction}

In computational complexity theory, the complexity class of
nondeterministic polynomial --- also known as NP ---
\cite{ladner:75,karp:72,cook:71,lucas:14} is one of the most fundamental
ones.  The class consists of decision problems that are {\em verifiable}
in polynomial time, however, no statement is made about the worst-case
complexity. Typically, the worst-case complexity scales in a
super-polynomial manner. The NP class includes a variety of notoriously
hard yet important optimization problems such as Ising spin glasses
\cite{barahona:82,katzgraber:14-ea}, the Boolean satisfiability problem
\cite{cook:71,karp:72,johnson:74}, minimum vertex covers
\cite{koenig:31,karp:72}, as well as the travelling salesman problem
\cite{applegate:07,arora:98}.  The aforementioned problems, as well as
many others in the NP class, have complex energy (cost function)
landscapes with many local minima and are typically only solvable in
super-polynomial (e.g., exponential or stretched exponential) times.
While at the moment efficient optimization algorithms cannot change the
worst-case (or even typical) complexity to a polynomial in the size of
the input, one can hope to at least have algorithms that have smaller
prefactors in the time complexity or stretched exponentials with smaller
exponents \cite{lucas:14,denchev:15,mandra:16}. This could result in
substantial speedup and allow for the study of problems with a
considerably larger number of variables. In fact, despite Moore's law
\cite{moore:65} hopefully still holding for the next few decades, large
advances can only be achieved by better algorithms and not raw computing
power.

Many optimization problems in the NP complexity class can be solved by
local search (LS) heuristics. This type of algorithm starts from a
candidate solution and then iteratively moves to a neighboring solution
with random or greedy moves of single Boolean variables. However, either
the greedy single-variable dynamics is quickly trapped in local minima
of the cost function, or exhaustively explores plateaus in the landscape
where no local moves can decrease the cost in a reasonable amount of
time.  To escape this single-move traps, randomizing moves can be
performed at the cost of additional computational time.  Paradigmatic
examples of (stochastic) local search algorithms have evolved from
algorithms such as GSAT and WalkSAT \cite{selman:93} for the maximum
satisfiability problem, NuMVC \cite{cai:13} for minimum vertex covers,
as well as simulated annealing and 2-opt algorithms
\cite{kirkpatrick:83,johnson:97} for the traveling salesman problem. For
spin glasses, methods such as extremal optimization \cite{boettcher:01},
local genetic algorithms \cite{pal:95} or the cluster-exact
approximation method \cite{hartmann:01,hartmann:04} have been
successful in tackling problems with up to approximately $2^{12}$
variables. In contrast to these local search algorithms that rely on
updating one variable at a time and, when trapped in a local minimum are
restarted from a new initial configuration, global update algorithms
flip multiple variables simultaneously in one iteration. This could, in
principle, lead to a large rearrangement of the variables and therefore
the ability to escape a local minimum.  Notable examples include the
building-block wise crossover in genetic algorithms \cite{whitley:94}
that facilitates inheritance of characteristics by an offspring from its
parents and global Swendsen-Wang \cite{swendsen:87} and Wolff cluster
algorithms \cite{wolff:89} for the simulation of ferromagnetic Ising
models in physics.  The latter are typically used to improve
thermalization for finite-temperature measurements and greatly reduce
autocorrelation times, thus massively speeding up the simulation and
allowing for a study of considerably larger system sizes.
Therefore, combining LS algorithms with carefully-designed nonlocal cluster
updates could allow for a wide-spread sampling of the phase space, and
hitherto speeding up the optimization of NP optimization problems.

In this work we design a stochastic search algorithm (\bor) that can
efficiently overcome local minima, as well as globally sample the cost
function landscape by large rearrangements of the variables.  To
overcome energy barriers efficiently, we combine the isoenergetic
cluster moves \cite{zhu:15b} with parallel tempering Monte Carlo.

Parallel tempering (PT) Monte Carlo
\cite{geyer:91,hukushima:96,katzgraber:09e} (also known as replica
exchange Markov chain Monte Carlo sampling) is a global update algorithm
aimed at improving the dynamic properties of Monte Carlo simulations of
physical systems. Although in this algorithm no clusters are generated
to perform large-scale variable rearrangements, in PT multiple copies of
the system with different initial conditions (i.e., independent Markov
chains) are simulated at different temperatures.  Then, based on the
Metropolis update criterion, one exchanges configurations at different
temperatures. This means that different copies of the system perform a
random walk in temperature space (in addition to the random walk in cost
function space from the single-variable Monte Carlo  updates) thus
allowing trapped configurations to be ``heated'' and then ``cooled''
with the goal of overcoming energy barriers and therefore escaping local
minima.  PT Monte Carlo has proven to be a versatile ``workhorse'' in
many research fields \cite{earl:05}, such as physics, biology and
chemistry.  Most importantly, by setting the lowest temperature of the
simulation to be close to zero, PT can used as an efficient optimizer
for any problem that can be cast into Boolean variables
\cite{moreno:03}.

The isoenergetic cluster move (ICM) algorithm \cite{zhu:15b}, related to
Houdayer's cluster updates \cite{houdayer:01}, requires two copies of
the Boolean system to be studied. Using information from the variables
in both systems, clusters are built in the dot-product space of the
variables. This cluster ``mask'' is then applied to both systems. As
shown in Ref.~\cite{houdayer:01}, the algorithm is rejection free, which
means that every update is accepted with probability $1$. Furthermore,
the value of the cost function (energy) of the combined system does not
change in the cluster move, which means that the two systems are
``teleported'' across phase space at a fixed value of the cost.
Typically, the cluster moves are combined with another host algorithm
(e.g., PT). The added large global rearrangements vastly improve the
overall performance of the host algorithm. This method has been used
extremely successfully in recent studies of spin-glass systems. See, for
example, Refs.~\cite{katzgraber:15,zhu:16,mandra:16}.

Fortunately, there is a close relationship between the statistical
physics of Ising spin glasses and a wide variety of Boolean NP problems
\cite{lucas:14}.  Mathematically, because the decision form of the Ising
spin glass model is NP-complete \cite{barahona:82}, there exists a
polynomial time mapping to any other NP-complete problem with Boolean
variables \cite{karp:72}. Here we demonstrate that \bor~--- a
combination of PT with ICM --- can be efficiently applied to a variety
of NP optimization problems, provided the cost function (Hamiltonian)
can be written as a polynomial function of a set of $N$ Boolean
variables $x_i$, i.e.,
\begin{equation}
- {\mathcal H}\left(x_1,\ldots, x_N\right) = 
  \sum_{i} h_i x_i 
+ \sum_{ij} J_{ij} x_i x_j 
+ \sum_{ijk} T_{ijk} x_i x_j x_k 
+ \ldots \,\,\, .
\label{eq:ham}
\end{equation}
We demonstrate the efficiency of the heuristic \bor~on the maximum
satisfiability and minimum vertex cover problems. Furthermore, we
compare the heuristic to current state-of-the-art heuristics, such as
CCLS \cite{luo:15}, DistUP \cite{DistUP:15}, Dist1 \cite{cai:14} and
NuMVC \cite{cai:13}.

The paper is organized as follows. In Sec.~\ref{sec:prelim} we introduce
the optimization problems used to illustrate \bor, followed by a detailed
description of \bor, parallel tempering Monte Carlo and the isoenergetic
cluster move algorithm in Sec.~\ref{sec:algo}.  Section \ref{sec:exp}
shows results on the Boolean satisfiability problem, as well as minimum
vertex covers, followed by concluding remarks.

\section{Studied benchmarks}
\label{sec:prelim}

In this section we briefly outline the benchmark optimization problems
used to illustrate the performance of the \bor~algorithm. Note that the
method can be applied to other difficult optimization problems, such as
spin glasses \cite{mandra:16}.

\subsection{Boolean satisfiability problem (SAT)}

The maximum satisfiability problem (MAX-SAT) is the combinatorial
optimization problem of determining a set of Boolean variables
$\{x_1,\ldots, x_N\}$ that maximize the number of satisfied clauses
$\{C_1,\ldots, C_M\}$ in a conjunctive normal form $\Psi = C_1\wedge
C_2\cdots \wedge C_M$, where
\begin{equation}
C_i = x_{i_1}\vee \ldots \vee x_{i_k}, \,\,\,\,\,\,\,\,\,\, 1\le i\le M .
\label{eq:ci}
\end{equation}
The variables $x_{i_1}, \ldots, x_{i_k}$ in Eq.~\eqref{eq:ci} are
selected from another set of Boolean variables, $x_1, \ldots, x_N,
\overline{x}_1,\ldots, \overline{x}_N$ with the goal of satisfying the
Boolean formula.  The weighted partial MAX-SAT problem is a
generalization of the maximum satisfiability problem in which each
clause $C_i$ is assigned a positive weight $w_i$. The objective of this
problem is to maximize the sum of weights of satisfied clauses by any
variable assignment. Note, also, that the partial MAX-SAT problem tries
to find an optimal assignment to the variables which satisfies all the
hard clauses and maximizes the number of soft clauses. The combination
of both variations is called the weighted partial MAX-SAT problem.
Weighted partial MAX-SAT problems are crucial elements of a broad range
in application areas such as telecommunications \cite{koster:99},
scheduling \cite{cha:97}, combinatorial online auctions \cite{fu:06}, as
well as circuit design \cite{marques:08}, to name a few.

Not-All-Equal Maximum Satisfiability (NAE-MAX-SAT) is one of the central
problems in theoretical computer science and is similar to MAX-SAT,
except for the additional requirement that at least one of the literals
in each clause be true and one be false.  NAE-MAX-SAT is symmetric with
respect to switching the Boolean variables \cite{krzakala:15}.  This
means that a representation of the form presented in Eq.~\eqref{eq:ham}
only has term with even powers of $x_i$.  The average Hamming distance
of the set of solutions is approximately $50\%$ of the total number of
variables and all solutions are statistically uncorrelated
\cite{coja:13}. This feature can be exploited to efficiently construct
probabilistic membership filters \cite{weaver:14} based on SAT formulas
\cite{fang:16}. A special case of NAE-MAX-SAT is the weighted
XOR-MAX-SAT problem where each clause contains XOR (exclusive or) rather
than an OR operators. Generally speaking, local search algorithms take
exponential time on random XOR-SAT formulas because flipping any
variable will dissatisfy all the currently satisfied clauses
\cite{ricci:01,jia:04}.

A Hamiltonian ${\mathcal H}$ (cost function) to describe MAX-SAT,
NAE-MAX-SAT, or weighted XOR-MAX-SAT may be written such that the
Hamiltonian is a measure of the number of unsatisfied clauses, i.e.,
${\mathcal H} =\sum_i w_i\overline{C}_i$.  The ground state(s) of this
Hamiltonian correspond to those assignments with all Boolean variables
that violate the minimum number of clauses.

\subsection{Minimum vertex cover problem}

A minimum vertex cover (MVC) is a vertex covering of a graph $G$ using
the smallest possible number of vertices. A graph $G=(V,E)$ is a set of
vertices $V$ and a set of edges $E$. A vertex cover of a graph $G$ can
simply be thought of as a set $S$ of vertices of $G$ such that every
edge of $G$ has at least one of member of $S$ as an endpoint.  Finding a
{\em minimum} vertex cover of a general graph is an NP-complete problem,
the complement of the maximum independent set problem \cite{koenig:31}.
The MVC problem has many real-world applications such as network
security, scheduling and industrial machine assignment \cite{gomes:06}.

Let $x_i$ be a Boolean variable on each vertex, which is $1$ if it is
colored, and $0$ if it is not colored. The Hamiltonian (cost function)
for MVC is given by ${\mathcal H} ={\mathcal H}_A + {\mathcal H}_B$.
The penalty term ${\mathcal H}_A$ imposes the constraint that every edge
has at least one colored vertex, i.e., 
\begin{equation}
{\mathcal H}_A = A\sum_{ij\in E}(1-x_i)(1-x_j) .
\end{equation}
Minimizing the number of colored vertices can be done by setting
\begin{equation}
{\mathcal H}_B = B \sum_i x_i .
\end{equation}
Choosing the coefficient $B < A$ with $A$ large ensures that it is never
favorable to violate the constraints imposed by ${\mathcal H}_A$.

\section{Outline of the \bor~algorithm}
\label{sec:algo}

Based on the ideas of reweighting hard constraints, parallel tempering
updates and isoenergetic cluster updates, we develop an efficient global
update algorithm \bor~for solving NP problems, which is outlined as
follows:
\begin{algorithm}
\caption{\bor}\label{alg:borealis}
\textbf{Input:} MAX-SAT instance, $maxMCS$ \\ 
\begin{algorithmic}[1]
\State {Re-weight hard clauses;}
\State {Initialize systems with random truth assignments;}
\For{$MCS=1$ to $maxMCS$}
\State {Metropolis update;}
\State {Parallel tempering update;}
\If{site percolation-threshold $p_c$ is high}
\State {Isoenergetic cluster update;}
\EndIf
\State {Keep track of lowest energy $E_{min}$ of all systems;}
\EndFor
\State \textbf{return} $E_{min}$
\end{algorithmic}
\end{algorithm}

We now describe the different updates performed in the \bor~algorithm in
detail.

\subsection{Weighting scheme in partial MAX-SAT and weighted partial
MAX-SAT}

Typically hard clauses have to be satisfied and satisfaction of soft
clauses is desirable but not mandatory. The simplest way to represent
the relative importance of hard clauses is to set their weights to the
number of soft clauses plus $1$. However, large weights on hard clauses
create large energy barriers and significantly slow down the search in
configuration space. An optimum weighting strategy adds weights to
constraints without distorting the solution space. In principle, one
could set the weights of all hard clauses to the number of soft clauses
left unsatisfied in an optimal solution (unfortunately, this not known).
Cha {\em et al.} \cite{cha:97} set weights to a hand-tuned optimal level
and Thornton \& Sattar \cite{thornton:98} later introduced two dynamic
constraint weighting schemes according to feedback received during the
search. Here, for the simplicity, for each hard clause we set the
weights to the maximum sum of the number of literals appearing in soft
clauses.

\subsection{Parallel tempering update}

Parallel tempering (PT) \cite{hukushima:96} is a simulation method aimed
at improving the dynamic properties of simple Markov chain Monte Carlo
simulations of physical systems. Essentially, $N_T$ copies of the
system, each with different initial conditions, are simulated at a range
of temperatures \{$T_1,T_2,...,T_{N_T}$\}. After a simple Monte Carlo
sweep of each variable of each copy of the system, configurations at
adjacent different temperatures are exchanged based on a Metropolis
criterion
\begin{equation}
p(E_i,T_i\rightarrow E_{i+1}, T_{i+1})=\min\{1, \exp(\Delta E \Delta \beta)\},
\label{:pt}
\end{equation}
where $\Delta \beta = 1/T_{i+1}-1/T_i$ is the difference between the
inverse temperatures and $\Delta E = E_{i+1}-E_i$ is the difference in
the energy of the two neighboring copies at different temperatures.  The
idea behind this method is to make configurations at high temperatures
available to the simulation at low temperatures, and vice versa. This
results in a very robust ensemble that is able to sample both low- and
high-energy configurations and easily overcomes energy barriers.

One important aspect of PT is that optimal temperature intervals must be
carefully chosen \cite{katzgraber:09e}.  When the temperatures are too
far apart, the energy distributions at the individual temperatures do
not overlap enough and many moves are rejected. If the temperatures are
too close, CPU time is wasted.  Although there are a wide variety of
``optimal'' approaches to determine the location of the individual
temperatures
\cite{predescu:04,kofke:04,kone:05,predescu:05,rathore:05,earl:05,katzgraber:06a,katzgraber:09e,machta:09,hamze:10},
usually the most accurate data for a fixed amount of computation are
obtained if we ensure that the acceptance probabilities for the
individual swaps between neighboring temperatures are approximately
independent of the temperature and roughly between $20\%-80\%$
\cite{rathore:05}. Any additional optimization of the temperatures
constitutes wasted efforts.

PT is an extremely powerful algorithm in the study of spin glasses. For
example, the speedup over conventional simple Monte Carlo for a cubic
spin glass with $N = 64$ variables at low temperatures is approximately
four orders of magnitude. This speedup grows with decreasing temperature
and an increasing number of variables $N$.

\subsection{Isoenergetic cluster moves}

We begin by simulating two copies of the system at multiple
temperatures. The cluster moves alone are not ergodic, as such, these
must be combined with simple Monte Carlo updates. One simulation step
using isoenergetic cluster moves (ICM) consists of the following steps:
\begin{enumerate}

\item{Identify a cluster in overlap space: Two independent
configurations (copies) are simulated at the same temperature. The site
overlap between copies $(1)$ and $(2)$, $q_i = x_i^{(1)} \oplus
x_i^{(2)}$, is calculated. This creates two domains in $q$-space: Sites
with $q_i = 0$ and $q_i = 1$.  Clusters are defined as the connected
parts of these domains in $q$-space \cite{comment:clusters}. One then
randomly chooses one site with $q_i = 1$ and builds the cluster by
adding all the connected variables with $q_i = 1$ in the domain with
probability $1$.  When no more variables can be added to the cluster in
$q$-space, the variables in {\em both} replicas that correspond to
cluster members in $q$-space are flipped with probability $1$,
irrespective of their value.}

\item{Perform one isoenergetic cluster move for all temperatures $T
\lesssim J$. Note that, in most cases, the characteristic energy scale
of the problem in Eq.~\eqref{eq:ham} is $J = 1$. This means that we
perform the cluster moves typically for $T \lesssim 1$.}

\end{enumerate}
In what follows we prove that ICMs leave the total energy of the
combined system of copies $(1)$ and $(2)$ intact.  Assume we randomly
pick a cluster $O^{\alpha}$ and the interaction matrices (or tensors)
associated with Boolean variables in $O^{\alpha}$ are represented as
$T^{\alpha}_{ijk\ldots}$.  $T^{\alpha}_{ijk\ldots}$ are comprised of two
different categories. The ones whose endpoints are all in the cluster
$O^{\alpha}$ and rest with only some endpoints in the cluster. Flipping
all the Boolean variables in the cluster $O^{\alpha}$ does not change
the total energy associated with interaction matrices in the first
category because
\begin{equation}
-\sum_{ijk\ldots} T^{\alpha}_{ijk\ldots} (x^{\alpha_1}_i x^{\alpha_1}_j
x^{\alpha_1}_k \ldots +x^{\alpha_2}_i x^{\alpha_2}_j x^{\alpha_2}_k \ldots)
\end{equation}
remains the same if we merely swap Boolean variables $x^{\alpha_1}_i,
x^{\alpha_1}_j, x^{\alpha_1}_k$ with $x^{\alpha_2}_i, x^{\alpha_2}_j,
x^{\alpha_2}_k$. For the interaction matrices in the second category,
the energy associated with these matrices is
\begin{equation}
-\sum_{ijk\ldots} T^{\alpha}_{ijk\ldots} (x^{\alpha_1}_i x^{\alpha_1}_j 
x^{\alpha_1}_k \ldots +x^{\alpha_2}_i x^{\alpha_2}_j x^{\alpha_2}_k \ldots) =  
-\sum_{ijk\ldots} T^{\alpha}_{ijk\ldots }x^{\alpha}_i (x^{\alpha_1}_j 
x^{\alpha_1}_k \ldots +x^{\alpha_2}_j x^{\alpha_2}_k \ldots) , 
\end{equation}
where the Boolean variables $x_j, x_k$ are included in the cluster
$O^{\alpha}$ whereas $x_i$ is not. Therefore, flipping Boolean variables
in the cluster only swap $x^{\alpha_1}_j, x^{\alpha_1}_k$ with
$x^{\alpha_2}_j, x^{\alpha_2}_k$ and leave the energy unchanged.  Note
that if the clusters in $q$-space percolate (i.e., extend the whole size
of the problem) a cluster update merely exchanges both configurations
and thus represent numerical overhead.  Generally speaking, the
performance of isoenergetic cluster moves is limited by the
site-percolation threshold $p_c$ of the topology of the underlying
problems \cite{comment:perc}, the amount of frustration present in the
system (that slows cluster growth as a function of temperature), as well
as the performance of vanilla parallel tempering.  For cases where $p_c$
is very small ($p_c \to 0$), the cluster updated can be removed from the
algorithm because they constitute unnecessary overhead.  In this case,
it is more efficient to simply run \bor~without the cluster updates. In
the next section we demonstrate how \bor~outperforms, for example, CCLS
on planar NAE-MAX-3SAT problems, random weighted XOR-MAX-2SAT, as well 
as NuMVC on minimum vertex covers with cluster moves included.

\section{Experiments}
\label{sec:exp}

\subsection{Benchmark problems studied}

For our empirical studies, we evaluate \bor~on a broad range of
benchmarks, including unweighted MAX-SAT, partial MAX-SAT, weighted
partial MAX-SAT, NAE-MAX-SAT, weighted XOR-MAX-SAT and MVC.  The MAX-SAT
instances comprise the most wide-spread benchmark, including random
instances from the Tenth Max-SAT Evaluation in 2015
\cite{maxsat_2015:15}.  To perform a scaling analysis we use the
\texttt{makewff} generator \cite{makewff:16}  with minor modifications
to generate random MAX-$k$-SAT ($M/N=30$) instances and weighted
XOR-MAX-2SAT instances ($M/N=1$) with certain clause-to-variable ratios.
The MAX-NAE-SAT and MVC instances used for our experiments are derived
from triangular lattices and random graphs with a ratio of edges to
vertices equal to $1.5$ (randomly selected). For planar NAE-MAX-3SAT
instances, literals in each clause are selected from variables with
random signs in a random triangle; the triangular lattice is known to
have a site percolation-threshold of $p_c=0.5$ which is optimal for
\bor.  For MVC, given a random graph of $N$ vertices and $M$ edges,
there exists a site percolation-threshold $p_c=N/2M=0.33$
\cite{bollobas:01} below which the network becomes fragmented while
above $p_c$ a giant connected component exists.  Coefficients $A$ and
$B$ in the MVC problem are chosen to be $1.3$ and $1$, respectively,
such that it is never favorable to violate the constraints imposed by
the penalty term ${\mathcal H}_A$.

\subsection{State-of-the-art algorithms}

We have compared \bor~to four local search solvers: CCLS, DistUP, Dist1
and NuMVC. CCLS combines a configuration checking strategy with a random
walk and has won four categories in the incomplete track of the 2015
Max-SAT Evaluation. Dist is a local search algorithm with a clause
weighting scheme and variable selection strategy. It has won the
weighted partial random MAX-SAT incomplete track of the 2015 Max-SAT
Evaluation. DistUP combines an assigning procedure PrioUP with the
solver Dist and has won the partial random Max-SAT incomplete track of
the 2015 Max-SAT Evaluation. Finally, NuMVC proposes two-stage exchanges
and edge weighting strategies for MVC.  It is largely competitive on the
DIMACS benchmark \cite{DIMACS:16} and dramatically outperforms other
state-of-the-art heuristic solvers on all BHOSLIB instances
\cite{BHOSLIB:16}.

\subsection{Machine specifications}

All experiments are carried out on the compute nodes of the Lonestar-5
high-performance computing cluster at the Texas Advanced Computing
Center, using a Xeon E5-2690 v3 (Haswell) $2.6$ GHz CPU and 64 GB
DDR4-2133 memory. The time limit is set to be $300$ seconds for each
instance. We have implemented \bor~in the programming language C and
complied it with \texttt{gcc} with \texttt{-O2} optimization.

\subsection{Results}

We first compare the the performance of \bor~to CCLS, DistUP and Dist1
from the Tenth Max-SAT Evaluation (2015). Then we illustrate how the
inclusion of ICM substantially improves the performance over vanilla PT
on planar NAE-MAX-3SAT, random weighted XOR-MAX-2SAT  and MVC instances.
Simulation parameters used in the experiments with \bor~are shown in
Table \ref{tab:borealisparams}.

\begin{table}
\caption{Parameters of \bor~for the different experiments in unweighted
MAX-SAT, partial MAX-SAT, weighted partial MAX-SAT, NAE-MAX-SAT,
weighted XOR-MAX-SAT and MVC benchmark problems. For each instance
category simulated, we perform $maxMCS$ Monte Carlo sweeps (and
isoenergetic cluster moves) for each of the $2 N_T$ copies of the
system.  $T_{\rm min}$ [$T_{\rm max}$] is the lowest [highest]
temperature simulated, and $N_T$ is the total number of temperatures
used in the parallel tempering Monte Carlo method. At each temperature,
two copies of the system are needed for the ICM updates.  Note that
isoenergetic cluster moves only occur for the lowest $N_{\rm c}$
temperatures simulated. If the column for $N_T$ has no value in 
parentheses, the simulations were done without the inclusion of
ICM updates.
\label{tab:borealisparams}}
\begin{tabular*}{\columnwidth}{@{\extracolsep{\fill}}l l l l l r }
\hline
\hline
Track & category  & $T_{\rm min}$ & $T_{\rm max}$ & $N_T(N_c)$ & $maxMCS$ \\
\hline
unweighted & all & $0.05$ & $1.23$   & $25$ & $30000$ \\
partial &min2sat & $0.10$ & $0.50$   & $80$ & $30000$ \\
partial &min3sat & $0.10$ & $0.50$   & $80$ & $30000$ \\
partial &pmax2sat & $0.10$ & $2.05$   & $40$ & $30000$ \\
partial &pmax3sat & $0.10$ & $2.05$   & $40$ & $30000$ \\
weighted partial  &abrame& $0.10$ & $39.00$   & $40$ & $30000$  \\
weighted partial  &wmax2sat& $0.10$ & $39.00$   & $40$ & $30000$ \\
weighted partial  &wmax3sat& $0.10$ & $39.00$   & $40$ & $30000$ \\
weighted partial  &wpmax2sat& $0.10$ & $23.50$   & $40$ & $30000$ \\
weighted partial  &wpmax3sat& $0.10$ & $7.90$   & $40$ & $30000$ \\
weighted XOR-MAX-2SAT & all & $0.05$ & $0.44$   & $40(40)$ & $2^{20}$\\
NAE-MAX-3SAT & all & $0.05$ & $1.23$   & $25(20)$ & $2^{20}$ \\
MVC & all & $0.05$ & $0.50$   & $40(40)$ & $2^{20}$ \\
\hline
\hline
\end{tabular*}
\end{table}

\begin{figure}[htp]
\begin{center}
\includegraphics[width=0.7\columnwidth]{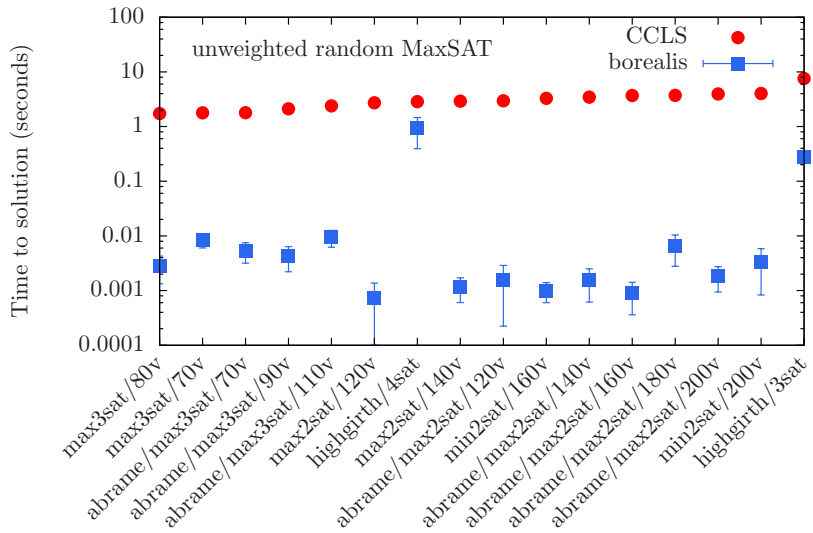}
\end{center}
\caption{Time to solution (CPU time) in seconds of \bor~and CCLS.  The
horizontal axis lists the different categories for the unweighted random
MAX-SAT track in the Tenth Max-SAT Evaluation (2015).  The time is
averaged over all instances in each instance category and error bars for
\bor~are computed via a bootstrap analysis. \bor~significantly
outperforms CCLS in all categories except for the highgirth instances
where it is still faster.}
\label{fig:unweighted_maxsat_2015}
\end{figure}

\begin{figure}[htp]
\begin{center}
\includegraphics[width=0.7\columnwidth]{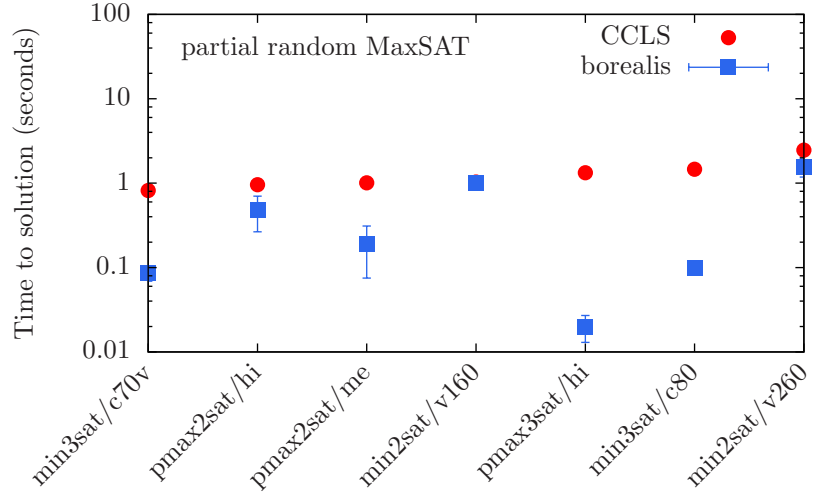}
\end{center}
\caption{Time to solution (CPU time) in seconds of \bor~and DistUP. The
horizontal axis lists the different categories for the partial random
MAX-SAT track in the 2015 Tenth Max-SAT Evaluation. \bor~significantly
outperforms DistUP in all categories except for the Min-2SAT instances
where the performance is comparable.}
\label{fig:partial_maxsat_2015}
\end{figure}

\begin{figure}[htp]
\begin{center}
\includegraphics[width=0.7\columnwidth]{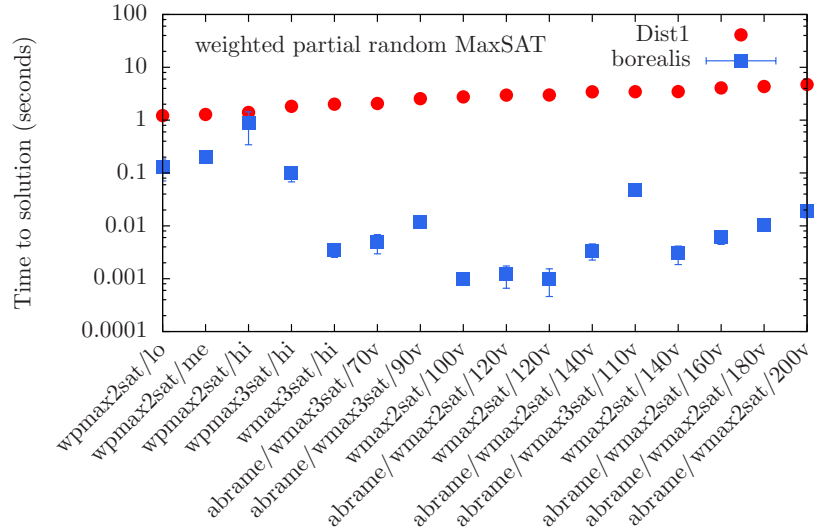}
\end{center}
\caption{Time to solution (CPU time) in seconds of \bor~and Dist1.  The
horizontal axis lists the different categories for the weighted partial
random MAX-SAT track in the Tenth Max-SAT Evaluation (2015). 
\bor~significantly outperforms Dist1 in all categories except for the
weighted partial MAX-2SAT instances with medium clauses to variables
ratio where the performance is comparable.}
\label{fig:weighted_partial_maxsat_2015}
\end{figure}

\begin{figure}[htp]
\begin{center}
\includegraphics[width=0.7\columnwidth]{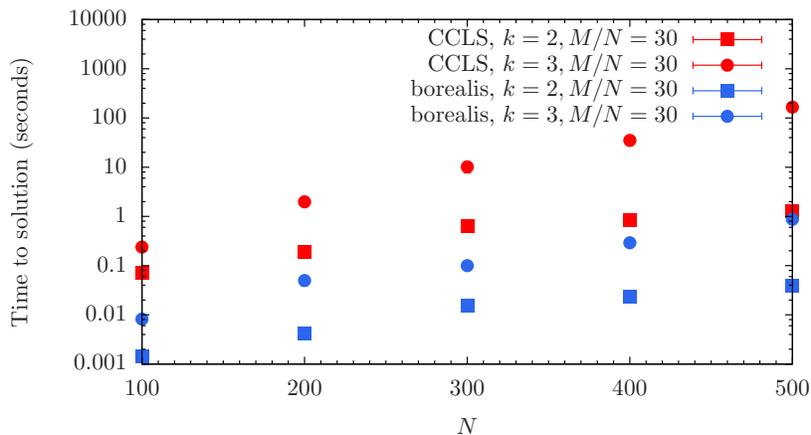}
\end{center}
\caption{Time to solution (CPU time) in seconds for \bor~and CCLS as a
function of system size $N$ for random unweighted MAX-SAT instances. The
time is averaged over $100$ instances for a given system size $N$. Error
bars are computed using a bootstrap analysis and are smaller than the
symbols. \bor~outperforms CCLS for all system sizes and scales better
with large $k$ and ratio $M/N=30$.}
\label{fig:unweighted_scaling}
\end{figure}

Figures \ref{fig:unweighted_maxsat_2015}, \ref{fig:partial_maxsat_2015},
and \ref{fig:weighted_partial_maxsat_2015} show the time to solution of
\bor~and CCLS, DistUP and Dist1 as a function category index for
unweighted, partial and weighted partial random MAX-SAT instances in
Tenth Max-SAT Evaluation (2015), respectively. \bor~(isoenergetic
cluster moves are not applied because most instances have a low
percolation threshold) finds solutions for all instances and
significantly outperforms CCLS, DistUP and Dist1 in most categories. In
the partial and weighted partial MAX-SAT benchmark instances, PT greatly
benefits from weighting schemes which lower energy barriers without
distorting the original solution space. Figure
\ref{fig:unweighted_scaling} demonstrates that \bor~scales better than
CCLS with large $k$ and ratio $M/N$.

\begin{figure}[htp]
\begin{center}
\includegraphics[width=0.7\columnwidth]{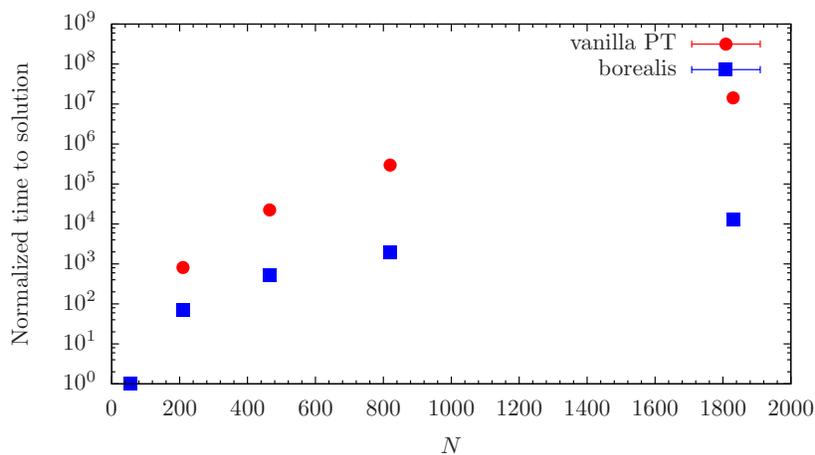}
\end{center}
\caption{Normalized time to solution (the time to solution for different
system sizes is divided by the time to solution for the smallest system
size) of \bor~compared to vanilla PT as a function of system size $N$
for planar NAE-MAX-SAT on a triangular lattice. The time is averaged
over $100$ instances for a given system size $N$. Again,
\bor~significantly outperforms vanilla PT and the speedup increases with
increasing system size.}
\label{fig:NAE_MAX_SAT}
\end{figure}

\begin{figure}[htp]
\begin{center}
\includegraphics[width=0.7\columnwidth]{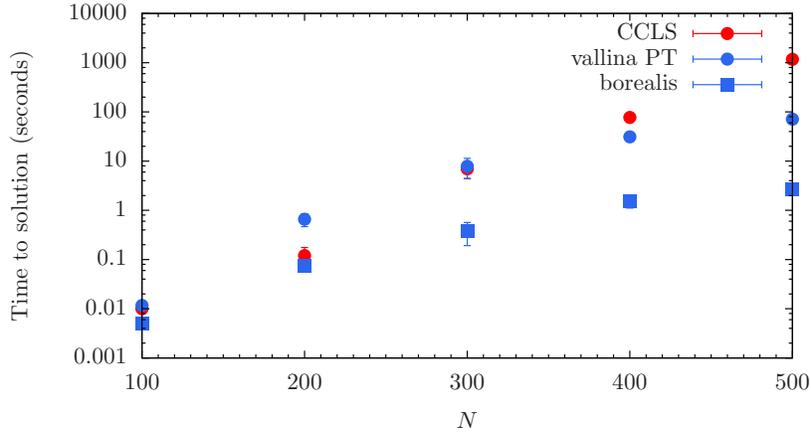}
\end{center}
\caption{Time to solution (CPU time) in seconds for \bor~(with/without
cluster moves) and CCLS as a function of system size $N$ for random
weighted XOR-MAX-2SAT instances ($M/N=1$). The time is averaged over
$100$ instances for a given system size $N$. Error bars are computed
using a bootstrap analysis and are smaller than the symbols. \bor~with
cluster updates outperforms CCLS on both CPU time and scaling.}
\label{fig:XOR_MAX_SAT}
\end{figure}

\begin{figure}[htp]
\begin{center}
\includegraphics[width=0.7\columnwidth]{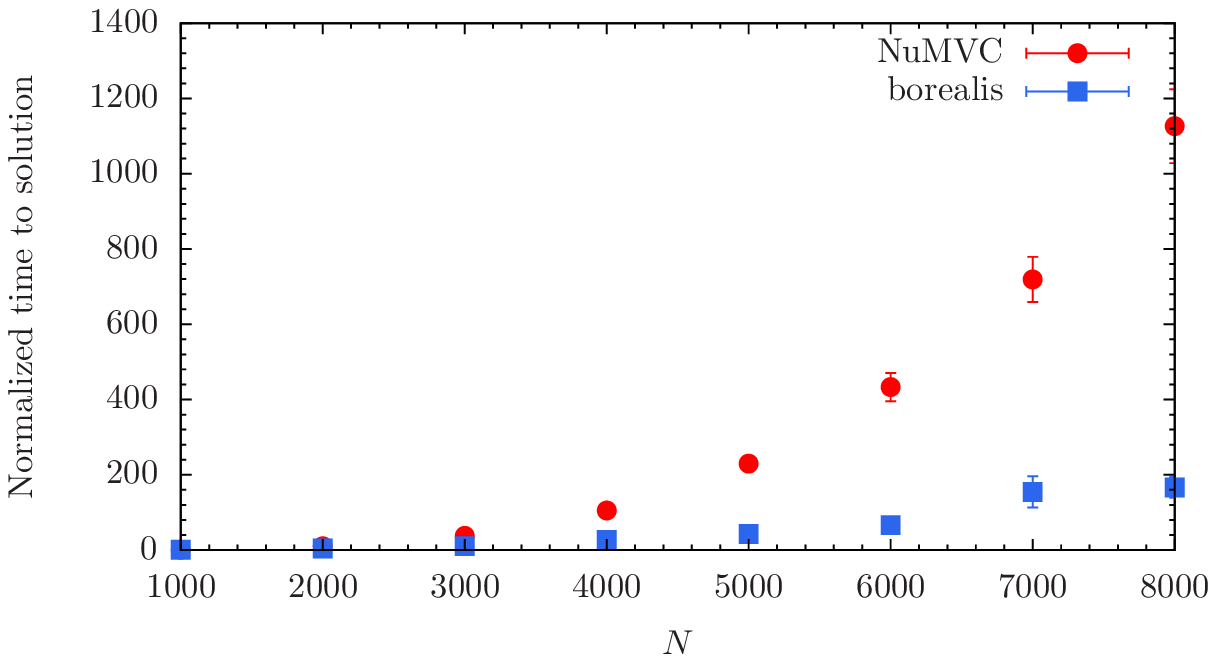}
\end{center}
\caption{Normalized time to solution (the time to solution for different
system sizes is divided by the time to solution for the smallest system
size) of \bor~and NuMVC as a function of system size $N$ for MVC on a
random graph with site-percolation threshold $p_c=N/2M=0.33$.
\bor~scales better than NuMVC and the speedup increases with increasing
system size. However, the advantage of \bor~over NuMVC is less
impressive due to the lack of frustrated interactions in MVC problems.}
\label{fig:MVC}
\end{figure}

Figure \ref{fig:NAE_MAX_SAT} shows the time to solution for \bor~and
vanilla PT as a function of system size $N$ for planar NAE-MAX-3SAT
instances. Each instance is run with and without isoenergetic cluster
moves for $10^6$ Monte Carlo sweeps. At the end of the runs we consider
solutions to be found if two results from each copy agree.  Clearly, the
inclusion of cluster moves results in a significant advantage over
vanilla PT for planar NAE-MAX-3SAT instances.  \bor~benefits from the
high site-percolation-threshold ($p_c=0.5$) of the triangular lattice,
as well as the relatively large cluster sizes due to the not-all-equal
constraint. Comparisons to the performance of CCLS on planar
NAE-MAX-3SAT are not included because CCLS has an extremely low success
rate in finding solutions for these instances. Figure
\ref{fig:XOR_MAX_SAT} shows the time to solution for \bor, vanilla PT
and CCLS as a function of system size $N$ for random weighted
XOR-MAX-2SAT instances. \bor~significantly outperforms CCLS and benefits
from the large rearrangement of the variables due to cluster moves and
tall energy barriers created by the XOR operators. In addition
to planar NAE-MAX-3SAT and weighted XOR-MAX-2SAT problems, we also
compare \bor~to the state-of-the-art local search algorithm NuMVC on the
MVC problems.  Figure \ref{fig:MVC} shows the time to solution of
\bor~and NuMVC as a function of system size $N$ for the MVC problem.
Again, \bor~clearly outperforms NuMVC on the scaling, however, its
advantage over vanilla PT is less impressive than for planar
NAE-MAX-3SAT problems due to the lack of frustrated interactions in MVC
problems.

\section{Conclusions}
\label{sec:concl}

We have developed a generic global update algorithm for NP optimization
problems with Boolean variables, \bor, based on a combination of
parallel tempering Monte Carlo \cite{hukushima:96} and isoenergetic
cluster moves \cite{zhu:15b}.  The global cluster moves, combined with
the tempering scheme allow for a wide-spread sampling of search space
and help escaping local minima separated by large energy barriers.  In
addition, by introducing a new weighting scheme in partial and weighted
partial MAX-SAT problems, we significantly lower the energy barriers
without distorting the solution space and show that \bor~outperforms
state-of-the-art algorithms on all random benchmark instances in the
Tenth Max-SAT Evaluation (2015).  For optimization problems with
relatively high site-percolation threshold, we demonstrate that the
inclusion of isoenergetic cluster moves significantly improves the
performance over vanilla parallel tempering on planar NAE-MAX-SAT
instances, random weighted XOR-MAX-2SAT and minimum vertex cover
problems on random graphs.

We intend to apply \bor~and cluster moves to other optimization problems
and algorithms, respectively, in the near future. For instance, cluster
moves can be added to speed up the study of fault diagnosis
\cite{perdomo:15b} in circuit design and genetic algorithms
\cite{hartmann:04} as a crossover operator to speed up the optimization.
In analogy to parallel tempering, we are also exploring the idea of
replica exchanges with different constraint strengths for
over-constrained NP optimization problems.

\section*{Acknowledgments}

We would like to thank 
Sergio Boixo,
Alexander Feldman,
Firas Hamze,
Bryan Jacobs,
Florent Krzakala,
Brad Lackey,
Jonathan Machta,
Oliver Melchert,
Hartmut Neven,
Andrew Ochoa,
Stefan Schnabel,
and
Martin Weigel
for fruitful discussions.
We would like to thank Karl F.~Roenigk for tasking us with this problem
and allowing us to explore it (successfully) in detail.  H.G.K.~would
like to thank Markus Blattner (Old Crow Inc.) for inspiration.  The
authors acknowledge support from the National Science Foundation (Grant
No.~DMR-1151387).  The work is supported in part by the Office of the
Director of National Intelligence (ODNI), Intelligence Advanced Research
Projects Activity (IARPA), via MIT Lincoln Laboratory Air Force Contract
No.~FA8721-05-C-0002. The views and conclusions contained herein are
those of the authors and should not be interpreted as necessarily
representing the official policies or endorsements, either expressed or
implied, of ODNI, IARPA, or the U.S.~Government. The U.S.~Government is
authorized to reproduce and distribute reprints for Governmental purpose
notwithstanding any copyright annotation thereon.  We thank Texas A\&M
University for access to their Ada and Curie clusters, as well as the
Texas Advanced Computing Center (TACC) at The University of Texas at
Austin for providing HPC resources (Lonestar Linux Cluster).

\vskip 0.2in

\bibliography{refs,comments}

\begin{thebibliography}{10}
\bibitem{comment:clusters}
Connected parts are determined by edges, which is straightforward for the
  vertex cover problem. For the Boolean satisfiability problem, we assume two
  different variables in any single clause are connected by an edge.
\bibitem{comment:perc}
In most cases, the determination of the site percolation threshold of the
  underlying graph relies on numerical approaches
  \cite{hoshen:76,newman:00,melchert:13,melchert:16} (e.g., Monte Carlo
  simulations combined with a finite-size scaling analysis). Significant
  algorithmic progress has been made by Newman and Ziff \cite{newman:00} for
  calculating the cluster size distribution or spanning probability over the
  entire range of site or bond occupation probabilities from zero to one in one
  single run (scaling linearly in the number of sites). In some cases --- such
  as random graphs \cite{bollobas:01} or triangular lattices \cite{skyes:63}
  --- it is even possible to derive the site percolation threshold
  analytically.
\end{thebibliography}

\end{document}